\documentclass[12pt]{iopart}
\usepackage{iopams}  
\usepackage{graphicx} \usepackage{times} \usepackage{color}
\usepackage{xspace} \usepackage{amsthm,amssymb}
\begin{document}

\title{Ergodic and non-ergodic anomalous diffusion in coupled stochastic processes}

\author{Golan Bel and Ilya Nemenman}

\address{Center for Nonlinear Studies
  and Computer, Computation and Statistical Sciences Division, Los
  Alamos National Laboratory, Los Alamos, NM 87545 USA}
\ead{golanbel@lanl.gov}

\begin{abstract}
Inspired by problems in biochemical kinetics, we study statistical
  properties of an over-damped Langevin process whose friction
  coefficient depends on the state of a similar, unobserved
  process. Integrating out the latter, we derive the long time
  behaviour of the mean square displacement. Anomalous diffusion is
  found. Since the diffusion exponent can not be predicted using a
  simple scaling argument, anomalous scaling appears as well. We also find that the coupling can lead to ergodic or non-ergodic behavior of the studied process. We
  compare our theoretical predictions with numerical simulations and
  find an excellent agreement. The findings caution against treating
  biochemical systems coupled with unobserved dynamical degrees of freedom by
  means of standard, diffusive Langevin descriptions.
\end{abstract}

\maketitle

\section{Introduction}
Single molecule kinetics has come within reach
of biophysical experiments \cite{Cluzel, English,Golding}, and
theoretical and computational tools for analysis of such processes
have experienced a corresponding growth \cite{Arkin, Paulsson,
  Nemenman, Wolynes, Sneppen}.  It is clear that the combinatorially
large number of microscopic steps involved in even the simplest of
biochemical events makes their rigorous stochastic treatment difficult. For
example, gene expression, often modeled as a single-step mRNA
creation, in fact, includes transcription-factor-DNA binding,
polymerase recruitment, transcriptional bubble formation, and multiple
elongation steps, each of which is a complex process on its own.

Therefore, any theoretical analysis of stochastic biochemical
processes necessarily involves coarse-graining: identifying a small
subset of dynamical variables that are modeled explicitly, while
agglomerating the rest into an effective behaviour. Such coarse-grained
dynamics is often modeled using the master equation or the Langevin
approaches, which require Markoviness or white-noise random forcing.
Both of these assumptions are, generally, flawed, and {\em
  quantitative} corrections have been worked out in certain cases
\cite{Szabo, Sinitsyn}. Less explored is the possibility that internal
degrees of freedom can introduce {\em qualitative} differences, such
as long-range temporal correlations among state transitions and
non-diffusive behaviour of the observed quantities.

A well-studied example shows that this is possible for a random walk
on a discrete lattice. In Ref.~\cite{WHcomb}, Weiss and Havlin
analyzed a two-dimensional diffusion model, known as the {\em comb}
model. There dynamics along the $y$ coordinate is unlimited, while
motion along $x$ is allowed only when $y=0$. This results in $\langle
x\rangle=0$ and $\langle x^2 \rangle\propto \sqrt{t}$, that is, in a
sub-diffusive motion of $x$. This model is hardly realistic in a
biochemical context due to the discontinuous dependence of the
diffusion coefficient on $y$. However, it is plausible that diffusive
dynamics of a real biological or chemical variable in the state space
or in the physical space depends on unobserved, decimated variables in
some other non-trivial way. For example, in a chemotaxing {\em E.\
  coli}, the number of unobserved signaling proteins {CheY-P} is
coupled to the distribution of times a bacterial motor rotates
counterclockwise, and the bacterium swims straight. For a fixed
concentration of {CheY-P}, obtained by modifying the chemotaxis
network, the distribution is essentially exponential \cite{cluzel-04},
resulting in a regular diffusive motion of the bacterium. But in a
wild-type bacterium, as the number of {CheY-P} fluctuates (diffuses in
the number space even for a constant external signal), the
distribution becomes a power law, and bacteria exhibit super-diffusive
real-space motion. While not true in this particular system, the
distribution of clockwise rotation times could have been strongly
coupled to {CheY-P} as well. This would have resulted in a power law
distribution of times that the bacterium spends reorienting itself
without moving forward, and hence in its sub-diffusive motion. In both
cases, neglecting the {CheY-P} fluctuations and describing bacterial
motion as a normal diffusion is qualitatively wrong.

In this paper, we abstract out the detailed biology and
explore these types of phenomena from the point of view of statistical
physics. We derive the properties of a diffusion process, for which the
diffusion coefficient depends on the state of another, unobserved, variable.
We show that, quite generally, such dependence
leads to anomalous diffusion of the observed process, suggesting that
traditional stochastic approaches may fail, and that more thought
should be given to modeling stochastic phenomena in complex
interacting systems, in particular in biophysics.

\section{The Model}
Our model is described by two variables $x$ and $y$, which may
represent, in particular, concentrations of two interacting chemical
species. The $x$ is considered as the observed quantity, while $y$ is
assumed hidden (i.e., unobserved). Particles
of both species can be created and destroyed, which results in an
over-damped diffusive motion of the system in the concentration space
(we disregard the directional drift for simplicity, but it can be
reintroduced easily). We assume that the diffusion of $x$ is
$y$-dependent. That is,
\begin{eqnarray}
\frac{dy}{dt}&=\frac{1}{\gamma_y}\eta\left(t\right), \label{ydyn} \\
\frac{dx}{dt}&=\frac{C\left(y\right)}{\gamma_x}\xi\left(t\right). \label{xdyn}
\end{eqnarray}
Here $\gamma_x,\gamma_y$ are the effective friction coefficients
(assumed to be homogeneous) corresponding to $x,y$;
$\eta\left(t\right)$ and $\xi\left(t\right)$ are independent,
zero-mean white noise forces such that
\begin{eqnarray}
  \langle\eta\left(t\right)\eta\left(t'\right)\rangle & =2 D_y \gamma_y^2 \delta\left(t-t'\right),  \label{WNCF1}\\
  \langle\xi\left(t\right)\xi\left(t'\right)\rangle & =2 D_x
  \gamma_x^2 \delta\left(t-t'\right). \label{WNCF2}
\end{eqnarray}
The idea of multiplicative noise was explored before. In \cite{Ihor}
the authors considered the case of a Langevin process (not
over-damped), in which a function of the velocity serves as a
``filter'' for the white noise. In \cite{Beck} the multiplicative
noise enters as a random friction coefficient; however, the
distribution of the random friction is independent of time. Similarly
in \cite{Ausloos} the noise enters as a random mass with distribution
which is again independent of time. Our model is different from the above mentioned models
since the distribution of the random coupling parameter $C(y)$ is time dependent;
moreover it depends in an arbitrary way (through the function $C(y)$) on
\emph{another} variable, which diffuses and hence has long range
temporal correlations. Further, the coupling considered in our model
does not introduce a directional bias since it is multiplied by a
white noise $\xi$, which takes both positive and negative values.  

The PDF of $y$ is that of a normal diffusion
\begin{equation}
p\left(y,t|y_0,0\right)=\frac{1}{\sqrt{4\pi D_y t}}e^{-\frac{(y-y_0)^2}{4D_yt}}, \label{ypdf}
\end{equation}
where the initial condition is $y(t=0)=y_0$.

The dynamics of the mean square displacement (MSD) $\langle
x\left(t\right)^2 \rangle$, where $\langle\cdots\rangle$ stands for
the average over the white noises $\eta$ and $\xi$, can be derived. Formally
integrating \Eref{xdyn} and substituting it in the expression for
the derivative of $x^2$ yields,
\begin{equation} 
\frac{dx\left(t\right)^{2}}{dt}=\frac{2C\left(y\right)}{\gamma_x^2} \xi\left(t\right)\int_{0}^{t} C\left(y\left(t'\right)\right)\xi\left(t'\right)dt'. \label{x2dyn}
\end{equation}
Averaging over the noise $\xi(t)$ yields the dynamics of the MSD of
$x$ conditional on $y(t)$,
\begin{equation}
\frac{d\langle x\left(t\right)^{2}|y(t)\rangle}{dt}=2 D_x C(y(t))^{2}. \label{ax2dyn}
\end{equation}
To get the marginal expectation $\langle x\left(t\right)^2\rangle$, we
now average the conditional expectation over $y$, which is distributed as in
\Eref{ypdf}:
\begin{equation}
\frac{d\overline{\langle x(t)^{2}\rangle}}{dt}=\frac{2D_x}{\sqrt{4\pi D_y t}}\int_{-\infty}^{\infty} e^{-\frac{(y-y_{0})^{2}}{4D_yt}}C(y(t))^{2}dy. \label{dax2}
\end{equation}

The function $C(y)$ may take different forms for different
systems. The simplest case is when the dynamics of $x$ is independent
of $y$ and $C\left(y\right)=C={\rm const}$.  Substituting this in
\Eref{dax2} yields the expected trivial result $\overline{\langle
  x^{2}\rangle}=2 D_x C^2 t$.

Another scenario is that $x$ can evolve in time only for a given range
of $y$ values ($|y|<y_{1}$), which resembles the discrete comb model
of Weiss and Havlin \cite{WHcomb}. Notice, however, that the comb
model is based on geometric constraints, i.e., the teeth, while in our
case the coupling between the two process is not due to the topology,
but rather to the physical nature of the processes. The similarity arises due the fact that in both models the first passage time distribution in the infinitely long $y$ axis, dominate the dynamics. Indeed,
substituting $C(y(t))=C
\Theta\left(y_{1}-y\left(t\right)\right)\Theta\left(y_{1}+y\left(t\right)\right)$,
where $C$ is a dimensionless constant, and $\Theta(y)$ is the
Heaviside Theta function, into \Eref{dax2}, we see that, for $t\gg
y_{1}^{2}/\left(12D_y\right)$, $\overline{\langle x\left(t\right)^2
  \rangle}\sim \sqrt{t}$ in agreement with \cite{WHcomb}. If $C(y)$
falls off exponentially as $y\to\infty$, the same sub-diffusion
exponent is recovered.

\begin{figure}[t]
\begin{center}
\includegraphics[width=\linewidth]{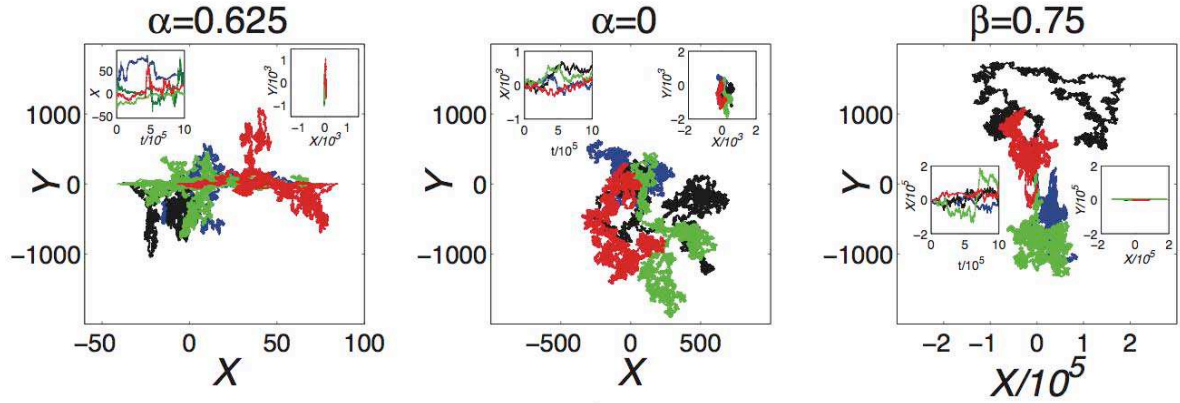}
\end{center}
\caption{Typical trajectories for different forms of the coupling function $C(y)$. We set $D_x=D_y=\gamma_x=\gamma_y=A=1$. In the left panel we used the repressive coupling of \Eref{CPL} with $\alpha=0.625$ to illustrate the suppression of
  diffusion in $x$. The central panel shows the case of $\alpha=0 $ in \Eref{CPL},
  namely decoupled Langevin processes resulting in normal diffusion. The right panel shows the case of enhanced diffusion in $x$ due to coupling of the form of \Eref{CSD} with $
  \beta=0.75 $. In each of the panels the left inset shows the observable $x$ as a function of time and the right inset shows $x$ and $y$ scaled equally. All trajectories start at $x=y=0$. The time step is set to $1$ and the total duration of each trajectory is $10^6$.}
\label{fig:1}
\end{figure}

A more interesting case is when, at large $y$, $C(y)$ falls off, but
not too sharply. We consider a power law form, namely
\begin{equation}
C(y(t))=\frac{1}{1+|Ay|^{\alpha}}, \label{CPL}
\end{equation}
where $A$ is a constant with the units of inverse length, and $\alpha$
is a dimensionless parameter. This is of a form of a repressive,
cooperative Hill kinetics, which describes many biochemical
processes \cite{albert08}. Typical diffusive trajectories with this $C(y)$, $A=1$, and
$\alpha=0,0.625$ are shown in Figure \ref{fig:1}.

Assuming that the behaviour of $C(y)$ for large $y$ (i.e., $C(y)\sim
|y|^{-\alpha}$) dominates the $t\to\infty$ dynamics of
$\overline{\langle x(t)^{2}\rangle} $, a simple scaling argument
suggests that $\overline{\langle x\left(t\right)^{2}\rangle}\sim
t^{1-\alpha}$. However, this is clearly wrong for large $\alpha$,
suggesting that $\overline{\langle x\left(t\right)^{2}\rangle}$ must
pick up an anomalous scaling due to the $y\to0$ properties of
$C(y)$. In what follows, we derive the long time behaviour of
$\overline{\langle x\left(t\right)^{2}\rangle}$ in a more rigorous
way.  

\Eref{dax2} with $C(y)$ as in \Eref{CPL}, and
considering the long time limit (${y_{0}}/{\sqrt{4 D_y t}}\ll1$)
gives
\begin{equation}
\fl \frac{d\overline{\langle x(t)^{2}\rangle}}{dt}=\frac{4D_x}{\sqrt{\pi}}\left\{\displaystyle{\int_{0}^{\frac{1}{A\sqrt{4D_yt}}}}\frac{e^{-y^{2}}}{\left[1+\left(\sqrt{4D_yt}Ay\right)^{\alpha}\right]^{2}}dy\right.
 \left.+{\displaystyle{\int_{\frac{1}{A\sqrt{4D_yt}}}^{\infty}}}\frac{e^{-y^{2}}}{\left[1+\left(\sqrt{4D_yt}Ay\right)^{\alpha}\right]^{2}}dy\right\}.\label{eq:1}
\end{equation}
In the first integral, we approximate the integrand as a constant for
$t\to\infty$, and, in the second integral, we neglect $1$ compared to
$\left(\sqrt{4D_yt}\, |Ay|\right)^{\alpha}$, thus obtaining
\begin{equation}
  \frac{d\overline{\langle x^{2}\rangle}}{dt} \approx \frac{2D_X}{A\sqrt{\pi D_yt}}+\frac{2D_xt^{-\alpha}}{\left(4A^2D_y\right)^{-\alpha}\sqrt{\pi}} \Gamma\left(\frac{1}{2}-\alpha,\frac{1}{4A^2D_yt}\right), \label{Dexp1}
\end{equation}
where $\Gamma\left(a,b\right)\equiv
{\int_{b}^{\infty}}\tau^{a-1}e^{-\tau}d\tau$ is the incomplete Gamma function.
Integrating \Eref{Dexp1} over $t$ results in the long time
behaviour of $\overline{\langle x(t)^{2}\rangle}$
\begin{equation}
\overline{\langle x(t)^{2}\rangle}\sim D_1 \sqrt{t}+D_2 t^{1-\alpha}, \label{x2diffexp}
\end{equation}
where $D_{1,2}$ are constants depending on the model parameters $D_x$,
$D_y$, $\alpha$ and $A$.  This implies that, for $\alpha<1/2$, the
long time behaviour is dominated by $\overline{\langle x(t)^{2}\rangle}
\sim t^{1-\alpha}$, as the scaling argument suggests. However, for
$\alpha>1/2$, the scaling argument breaks and $\overline{\langle
  x(t)^{2}\rangle} \sim \sqrt{t}$. Note that when the
$C\left(y\right)$ falls faster than $1/\sqrt{y}$, the diffusion
exponent is exactly the same as in the case in which the dynamics of
$x$ is limited to a finite range near $y=0$.

The case of $\alpha=\frac{1}{2}$ is special and the integral of \Eref{eq:1} can be calculated exactly, yielding
\begin{eqnarray}
&\frac{d\overline{\langle x(t)^{2}\rangle}}{dt} = G^{5 \ 4}_{4 \ 5} \left(\frac{1}{4A^2D_yt}\left|
\begin{array}{c}
-\frac{1}{4},0,\frac{1}{4},\frac{1}{2} \\
0,0,\frac{1}{4},\frac{1}{2},\frac{3}{4} 
\end{array} \right. \right)
\sim \frac{\ln t}{\sqrt{t}}, \label{ahalf}
\end{eqnarray}
where $G$ denotes the Meijer G function \cite{Mathematica}. The
leading order term of the MSD is then
\begin{equation}
\overline{\langle x(t)^{2}\rangle}\sim\sqrt{t}\ln t.
\end{equation}

\begin{figure}
\includegraphics[width=\linewidth]{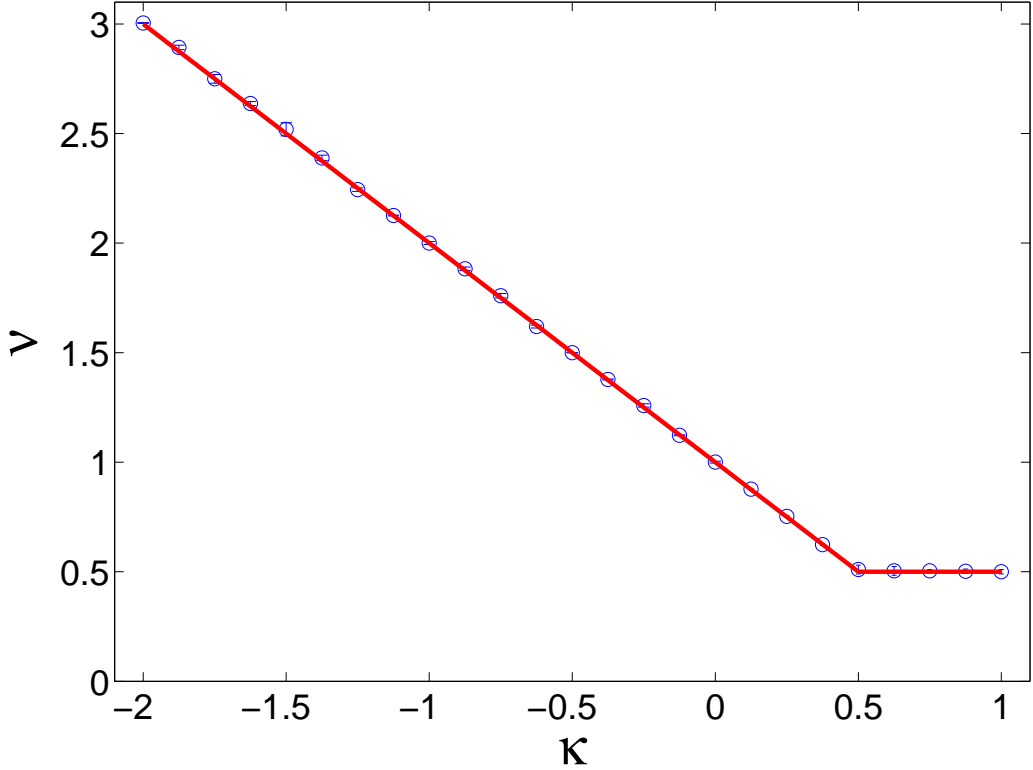}
\caption{ Leading order diffusion exponent $\nu$ defined by
  $\overline{\langle x(t)^{2}\rangle}\sim t^{\nu} $ for the coupled
  stochastic processes model with different couplings between the
  diffusion of $x$ and $y$, measured by the exponent $\kappa$. For the
  case of hindered diffusion of $x$, \Eref{CPL}, $\kappa=\alpha$,
  while for the enhanced diffusion, \Eref{CSD},
  $\kappa=-\beta$. Numerical simulations (points) and theoretical
  predictions (line) agree for both scenarios.}
\label{fig:2}
\end{figure}
The coupling function $C(y)$ depends only on the variable $y$ which
undergoes normal diffusion. Thus we can derive the probability
distribution of $C(y)$. In order to simplify the notation, in what
follows we use $C$ to denote the coupling function without explicitly
specifying its dependence on $y$. When $C$ is given by \Eref{CPL}, its
probability density is:
\begin{equation}
P\left(C\right)=\frac{2}{\alpha C^{\frac{\alpha+1}{\alpha}}\left(1-C\right)^{\frac{\alpha-1}{\alpha}}A\sqrt{4\pi D_{y}t}}e^{-\frac{\left(1-C\right)^{2/\alpha}}{4D_{y}tA^{2}C^{\frac{2}{\alpha}}}}, \label{PCsub}
\end{equation}
where $0 \leq C \leq 1$.  In the above derivation, we set $y(t=0)=0$
for simplicity. We see that the distribution of the noise introduced
by $C$ is time dependent. Note that the temporal correlations of $C$ are also important for the dynamic properties of $x$.

So far we have considered only situations in which the motion of $x$ was
slowed at large $y$, but we can also consider the opposite scenarios,
when large $y$ promotes diffusion in $x$, as in \cite{cluzel-04}:
\begin{equation}
C\left(y(t)\right)=|Ay|^{\beta},\quad \beta>0.  \label{CSD}
\end{equation}
This now resembles the Hill activation kinetics for low concentration
of the substrate molecules \cite{albert08}. Typical trajectories for this form of the
coupling function with $\beta=0.75$ are shown in Figure~\ref{fig:1}.

For coupling of this form, the dynamics of the MSD(\Eref{dax2}) yields
\begin{equation}
\frac{d\overline{\langle x(t)^{2}\rangle} }{dt}=\frac{2D_x}{\sqrt{\pi}}\left(4D_yt\right)^{\beta} \int_{-\infty}^{\infty}e^{-\left(y-\frac{y_{0}}{\sqrt{4D_yt}}\right)^{2}}|Ay|^{2\beta}dy, \label{SDx2}
\end{equation}
which, in the long time limit, gives ${d\overline{\langle
    x(t)^{2}\rangle}}/{dt}\sim t^{\beta}$, and
\begin{equation}
\overline{\langle x(t)^{2}\rangle}\sim t^{\beta+1}.
\end{equation}
The distribution of the coupling function $C$ in this case is given by
\begin{equation}
 P\left(C\right)=\frac{C^{-\frac{\beta -1}{\beta }} e^{-\frac{C^{2/\beta }}{4 t}}}{\sqrt{\pi } \sqrt{t} \beta }, \label{PCsuper}
\end{equation}
where $C\geq0$.

To confirm our analytical results, we performed numerical simulations
for the different cases considered above. In Figure~\ref{fig:2}, we
present a comparison of the diffusion exponent $\nu$ (defined by
$\overline{\langle x(t)^{2}\rangle}\sim t^{\nu} $) versus the coupling
parameter $\kappa$ (for the sub-diffusion scenario, \Eref{CPL},
$\kappa=\alpha$, and for the super-diffusion scenario, \Eref{CSD},
$\kappa=-\beta$) between the simulations and the analytical
results. The simulations where done according to
Eqs.~(\ref{ydyn},~\ref{xdyn}) with $\gamma_{x,y}=1$, $D_{x,y}=1$ and
$dt=1$. We averaged the results over $10^4$ trajectories, each of
duration $10^7\dots10^8 dt$.

A simple linear regression to $\log \overline{\langle
  x(t)^{2}\rangle}= \nu \log t +{\rm const}$ was performed to estimate
$\nu$.  Since the standard parameter errors obtained for the
regressions were negligible, the error bars of $\nu$ were estimated
from the variability of the fitted values as we changed the domain of
$t$, for which the fits were performed. Figure~\ref{fig:2} shows a clear
agreement between our theoretical results and the simulations.  Note
that, in certain cases, the convergence to the leading behaviour of
$\overline{\langle x(t)^{2}\rangle} $ as $t\to\infty$ is slow since
the difference between the exponents of the leading and the sub-leading
terms is small. This slowness determined the lengths of the
simulations. 

\section{Time Averaged Mean Square Displacement}
There are many models of anomalous diffusion, including a time
dependent friction coefficient in the Langevin equation \cite{Luczka},
continuous time random walk (CTRW) \cite{SherMont}, fractional
Brownian dynamics \cite{Mandelbrot}, fractional Langevin dynamics
\cite{Min,Goychuk,Burov} and Langevin dynamics with coloured noise
\cite{Muralidhar}, to name a few. For a new model resulting in an
anomalous diffusion, it is important to see if it can be reduced to
one of these more familiar constructions. For example, the
$t\to\infty$ behaviour of the original comb model is equivalent to
CTRW \cite{BouchaudRev} with a power-law tail of the distribution of
the times between successive jumps along $x$.  However, in our model,
the analogy is not as straightforward since the continuous dynamics of
$y$ induces temporal correlations among successive motions along $x$.

To understand the relation of the coupled diffusion model to the
others in the literature, we note that all of them yield the same
behaviour for the {\em ensemble} averaged MSD in the long time
limit. However, they still differ from each other in the short time
behaviour, the shape of the distribution, and even in the long time
behaviour of {\em time} averaged quantities (for example, the CTRW
exhibits ergodicity breaking \cite{Bel}). In particular, the time
averaged MSD (TAMSD) is an important property (it is the TAMSD that is
observed in typical single molecule diffusion experiments in
biological systems \cite{Golding,Tolic}, and the number of recorded
trajectories is often insufficient to estimate ensemble averages).

The TAMSD is defined as 
\begin{equation}
\overline{\delta^2} \left(\Delta,t \right) = \frac{1}{t-\Delta} \displaystyle{\int\limits_0^{t-\Delta}} \left[ x\left(\tau+\Delta\right)-x\left( \tau \right)\right]^2 d \tau, \label{TAMSD}
\end{equation}
which averages the squared displacement of a particle in time $\Delta$ (the time lag)
over a single particle trajectory of duration $t$.  For CTRW, the
TAMSD is a random quantity and even its ensemble average still
exhibits aging, that is, dependence on the measurement duration
\cite{Barkai,Klafter} $t$ in \Eref{TAMSD}. On the contrary, for the fractional Brownian
and Langevin dynamics, the TAMSD converges to the ensemble average MSD
for long times \cite{Barkai2}.

\begin{figure}
\includegraphics[width=\linewidth]{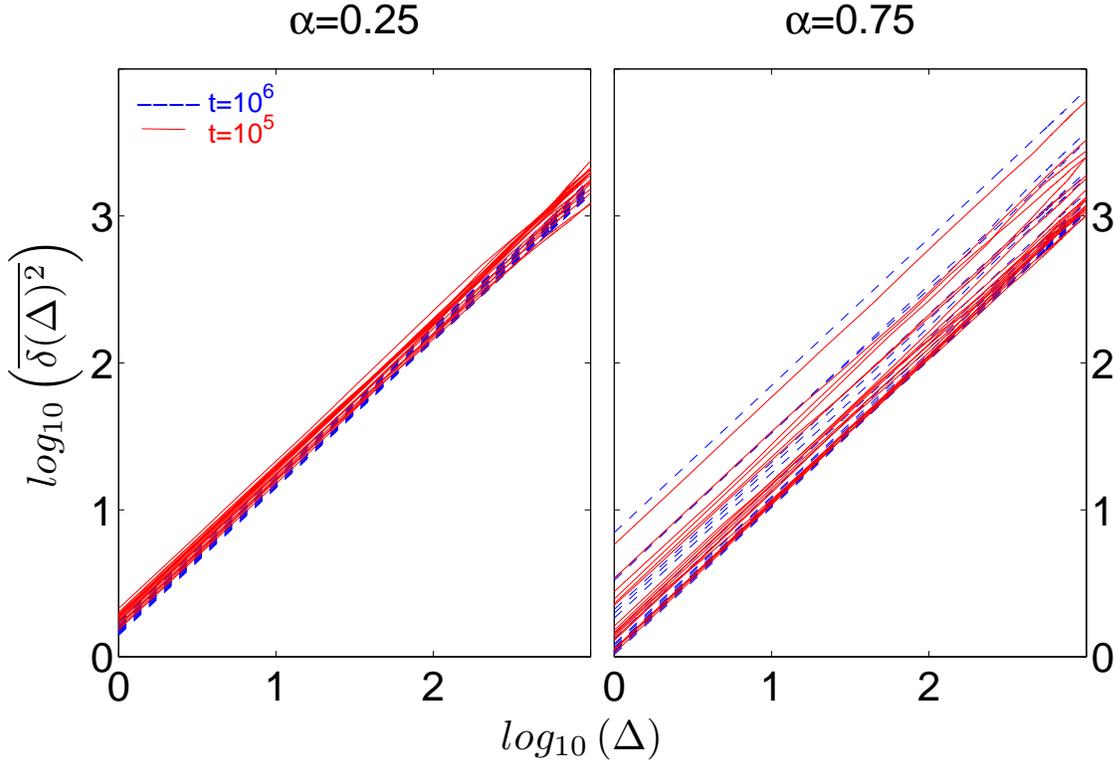}
\caption{ The time averaged MSD $\overline{\delta^2}\left(\Delta,t
  \right)$ versus $\Delta$ for $\alpha=0.25,0.75$ (left and right
  panels, respectively). All other parameters are the same as for the
  previous figures. We used 20 trajectories of duration $t=10^5$
  (solid red lines) and 20 trajectories of duration $t=10^6$ (dashed
  blue lines). The straight lines with slope $1$ in the log-log scale
  reflect a linear dependence of the TAMSD on the time lag
  $\Delta$. For $\alpha=0.25$, the TAMSD of different trajectories
  converges as the measurement time increases (the blue lines
  collapse, and the red do not), namely a time average of a single
  particle behaves like the ensemble average. However, for
  $\alpha=0.75$, there is no such convergence, just like for CTRW (the
  coefficient of proportionality between $\overline{\delta^2}$ and
  $\Delta$ varies from trajectory to trajectory). In both cases, the
  ensemble average of the TAMSD decreases as the measurement time
  increases, indicating aging.  }
\label{fig:3}
\end{figure}

We investigated the behaviour of the TAMSD in our model with
repressive coupling numerically. We find that, when the scaling
argument holds, namely for $\alpha \leq 1/2 $ (see \Eref{CPL}), the
TAMSD is not a random quantity, but it still shows aging, as we would
expect for Langevin dynamics with a time dependent friction. On the
other hand, when $\alpha \geq 1/2$, and the diffusion exponent is
$\nu=1/2$, the TAMSD shows a similar behaviour to that of the CTRW
\cite{Barkai}.  In Figure~\ref{fig:3}, we show the TAMSD versus
the time lag $\Delta$ for $\alpha=0.75$ and $\alpha=0.25$. Each line shows the
TAMSD for a single trajectory. The solid red lines represent
trajectories of duration $t=10^5$, and the dashed blue lines are for
$t=10^6$. All lines show a linear dependence on $\Delta$, but with
different coefficients. For $\alpha=0.25$, the TAMSD lines converge as
the trajectory duration grows (the dashed blue lines essentially
overlap), while for $\alpha=0.75$, the lines remain random. This is a
clear indication of ergodicity breaking in our model for
$\alpha>0.5$. Further, this analysis suggests that the coupled
Langevin processes model stands as its own class among other anomalous
diffusion models, exhibiting time-dependent diffusion coefficient
Langevin dynamics for certain forms of coupling, and some aspects of
ergodicity-breaking CTRW for others.

\section{Discussion} In this article, we introduced a model of coupled
diffusive processes, where the diffusion of the observed variable $x$
is coupled to the value of a hidden variable $y$. We showed that the
dynamics of $x$ exhibits anomalous diffusion for every considered form
of the coupling between the variables. Depending on the nature of the
coupling, the motion of $x$ can be sub- or super-diffusive (and even
super-ballistic, as is the case of a frictionless particle subject to
a white noise). Further, even for an arbitrary ``strong'' (such that
the dynamics of $x$ is limited to a small range of $y$ values)
repressive $xy$ coupling, the $x$ diffusion exponent $\nu$ is limited
from below by $\frac{1}{2}$ (anomalous scaling), so that localization
of $x$ is impossible. Even though the long-time ensemble-averaged
behaviour of our model is similar to that of many others describing
anomalous diffusion, the model does not reduce to any of them,
exhibiting an effective time-dependent diffusion coefficient, aging,
and ergodicity breaking for different values of its parameters.

The anomalous scaling and the ergodicity breaking appear for
``strong'' $xy$ coupling (i.e., large $\alpha$). This is because, for
$\alpha<1/2$, motion of particles away from $y=0$ contributes
substantially to the ensemble averaged MSD of $x$.  On the contrary,
for $\alpha>1/2$, only motion near $y=0$ is important, namely the
first passage time in an infinite one dimensional system (the
diffusion process in $y$) plays a major role in the dynamics of the
observed quantity, $x$. It was also verified numerically that, for
$\alpha>1/2$, where the dynamics of $x$ is dominated by motion in a
narrow range near $y=0$, the propagator is non-Gaussian as expected
from a CTRW-like renewal process. A similar result holds for the
super-diffusive regime, $\beta>0$.  On the other hand when the motion
at any $y$ is important ($0\le \alpha<1/2$), the process in $x$ has
long-time correlations and the propagator takes a Gaussian form
(similar to that of a diffusion process with a time dependent
diffusion coefficient).

While important in its own right, the coupled diffusion model raises
its most important questions in the biological domain.  Unobserved
dynamical quantities lead to anomalous diffusion in {\em E.\ coli}
chemotaxis \cite{cluzel-04}, or in {mRNA} diffusion in cells
\cite{Golding}. Further, some of the best established models of
cellular regulation involve coarse-graining of dynamics. For example,
in some models of the {\em E.\ coli} {\em lac} operon, the {\em lac}
repressor itself is an unobserved variable \cite{wall08}, which is
coupled to the speed of production of the lactose permease and the
lactose-utilizing enzyme and, through them, to the import and the
degradation of lactose in the cell. Since any coupling may lead to
anomalous diffusion and in some cases even to ergodicity breaking, it
begs the question whether relying on the common Langevin or master
equation analysis of stochasticity of the {\em lac} repressor or other
regulatory circuits, such as the $\lambda$-phage \cite{Arkin,Sneppen},
{\em mar} \cite{Cluzel}, and others, is rigorously justifiable.

\ack We thank E.\ Barkai and A.\ Zilman for stimulating discussions
and LANL Center for Nonlinear Studies for support and for striving to maintain an island of sanity in the increasingly more complex environment. This work was funded by LANL Laboratory Directed Research
and Development program.


\end{document}